# Physically Plausible Vectorial Metrics for Polarization Information Analysis

Runchen Zhang[*], Xuke Qiu, Yifei Ma, Zimo Zhao, An Aloysius Wang, Jinge Guo, Ji Qin, Steve J. Elston, Stephen M. Morris[*], and Chao He[*]

*Department of Engineering Science, University of Oxford, Parks Road, Oxford, OX1 3PJ, UK*

*Corresponding authors:* runchen.zhang@eng.ox.ac.uk; stephen.morris@eng.ox.ac.uk
chao.he@eng.ox.ac.uk

The Mueller Matrix Polar Decomposition method decomposes a Mueller matrix into a diattenuator, a retarder, and a depolarizer. Among these elements, the retarder, which plays a key role in medical and material characterization, is modelled as a circular retarder followed by a linear retarder when using this approach. However, this model may not accurately reflect the actual structure of the retarder in certain cases, as many practical retarders do not have a layered structure or consist of multiple (unknown) layers. Misinterpretation, therefore, may occur when the actual structure differs from the model. Here we circumvent this limitation by proposing to use a physically plausible parameter set that includes the axis orientation angle $\varphi$, the degree of ellipticity $\chi$, and the elliptical retardance $\rho$. By working with this set of parameters, an overall characterization of a retarder is provided, encompassing its full optical response without making any assumptions about the structure of the material. In this study, experiments were carried out on liquid crystalline samples to validate the feasibility of our approach, demonstrating that the physically plausible parameter set adopted provides a useful tool for a broader range of applications in both biomedical imaging and optical material analysis.

# 1. Introduction

Mueller matrix polarimetry, valued for its ability to reveal structural information of samples, is gaining prominence in vectorial optics and sample analysis. The 16 elements of a Mueller matrix comprehensively describe the polarization properties of a sample, which are crucially linked to the microstructure of the material. This renders it particularly useful in medical diagnostics[1–23] or material analysis[24–31]. However, due to its mathematical complexity and abstract nature, being able to extract meaningful physical features directly from the Mueller matrix remains a challenge. Towards this end, scientists have developed various methods to analyze the Mueller matrix, with typical examples including Mueller matrix symmetric decomposition (MMSD), Mueller matrix diffearential decomposition (MMDD)[32,33], Mueller matrix transformation (MMT)[34], and Mueller Matrix Polar Decomposition (MMPD)[35]. Among these methods, MMPD stands out as a prevalent method which has been used and validated in characterising various biomedical or optical materials, particularly through its retarder component[36–49].

The retarder component is conventionally characterized by the following three parameters: for the linear retarder, we have the linear retardance $\delta$ and the linear fast axis orientation $\theta$ whereas for the circular retard this is characterized by the circular retardance $\phi$[16]. It is noted that the handedness of the circular retarder is represented by the sign of the circular retardance. Previous studies have shown that the linear retarder strongly correlates with birefringent fibrous structures in biomedical tissues[8,50–55]. Specifically, the orientation of the linear axis can reveal disorganized fiber orientations, indicating symptoms of disease, while linear retardance measures the density of the fibrosis in various tissues[51–54,56,57]. These properties change markedly across different disease stages, making linear retarder parameters strongly correlated with the diagnosis and staging of diseases such as breast cancer, bowel disease, cervical cancer, and liver cancer[51,52,58–64]. Meanwhile, circular retardance primarily occurs in chiral molecules such as glucose, skin, and soft tissue membranes[65–67]. The magnitude and distribution of circular retardance provides insights into the composition and structure of the material[66,68–70].

Despite the obvious importance of these parameters, misinterpretations can occur when attempting to decipher the properties of the retarder using this approach. Specifically, for retarder samples with a two-layer structure, the conventional MMPD parameter set accurately represents the configuration only when it consists of a circular retarder followed by a linear retarder (see Section 2), due to reciprocity and predefined formats. Also, in practice, some retarders exhibit a multilayered structure, as seen in bulk tissue, and specific information about their hierarchical structure, such as the number of layers and the properties of each layer, is often unknown. In addition, some retarders do not have a distinct layered structure, making layer-based modelling inherently unsuitable. Without prior knowledge of the target, assuming that the retarder element is a fixed bilayer structure may yield unreliable results. Therefore, it is essential to adopt parameters that can reliably characterize retarder properties from a general perspective without prior knowledge of the actual structure.

This study focuses on appropriately describing general retarders in the context of MMPD. Firstly, the limitations of conventional parameters are highlighted. Then, a physically plausible set of vectorial metrics is proposed to describe the overall effect of the retarder from the perspective of an elliptical retarder, circumventing the ambiguities associated with further decomposition. Finally, the physical plausibility of this set of metrics is tested using nematic and chiral nematic phase liquid crystal (LC) samples, which serve as well-defined retarder models due to their stable and known birefringent properties.

## 2. Theory

### a) Background

MMPD decomposes a Mueller Matrix into a sequential multiplication of a diattenuator matrix ($M_D$), a retarder matrix ($M_R$), and a depolarizer matrix ($M_\Delta$) as shown in Eq. (1) [see further details in **Supplementary Material 1**].

$$M = M_\Delta M_R M_D \quad (1)$$

Conventionally, $M_R$ is further decomposed into the product of a circular retarder matrix $M_{CR}$ and a linear retarder matrix $M_{LR}$[71], as shown in Eq. (2). The circular retarder is characterized by the circular retardance $\phi$ rather than the optical rotation $\psi$ (to align with the characterization of the linear retarder), while the linear retarder is characterized by the linear retardance $\delta$ and the linear axis orientation $\theta$.

$$M_R = M_{LR} M_{CR} = \begin{pmatrix} 1 & 0 & 0 & 0 \\ 0 & \cos^2 2\theta + \sin^2 2\theta \cos\delta & \sin 2\theta \cos 2\theta (1-\cos\delta) & -\sin 2\theta \sin\delta \\ 0 & \sin 2\theta \cos 2\theta (1-\cos\delta) & \sin^2 2\theta + \cos^2 2\theta \cos\delta & \cos 2\theta \sin\delta \\ 0 & \sin 2\theta \sin\delta & -\cos 2\theta \sin\delta & \cos\delta \end{pmatrix} \begin{pmatrix} 1 & 0 & 0 & 0 \\ 0 & \cos\phi & \sin\phi & 0 \\ 0 & -\sin\phi & \cos\phi & 0 \\ 0 & 0 & 0 & 1 \end{pmatrix} \quad (2)$$

Notably, besides the aforementioned decomposition alignment, $M_R$ can also be decomposed in reverse order, as the product of a linear retarder $M_{LR}$, and a circular retarder $M_{CR}$, as shown in the following equation.

$$M_R = M_{CR} M_{LR} = \begin{pmatrix} 1 & 0 & 0 & 0 \\ 0 & \cos\phi & \sin\phi & 0 \\ 0 & -\sin\phi & \cos\phi & 0 \\ 0 & 0 & 0 & 1 \end{pmatrix} \begin{pmatrix} 1 & 0 & 0 & 0 \\ 0 & \cos^2 2\theta + \sin^2 2\theta \cos\delta & \sin 2\theta \cos 2\theta (1-\cos\delta) & -\sin 2\theta \sin\delta \\ 0 & \sin 2\theta \cos 2\theta (1-\cos\delta) & \sin^2 2\theta + \cos^2 2\theta \cos\delta & \cos 2\theta \sin\delta \\ 0 & \sin 2\theta \sin\delta & -\cos 2\theta \sin\delta & \cos\delta \end{pmatrix} \quad (3)$$

Since matrix multiplication is non-reciprocal, a natural question arises: will the multiplication sequence have an impact on the decomposition result, and if so, how significant will it be? Our analysis shows that reversing the multiplication sequence maintains the values of the circular retardance $\phi$ and linear retardance $\delta$, but alters the linear axis orientation $\theta$ from $\theta$ to $\theta + 0.5\phi$ [see details in **Supplementary Material 2**]. In certain biomedical samples, the linear axis orientation can indicate fiber orientation, so an incorrect value may lead to misinterpretation of tissue structure.

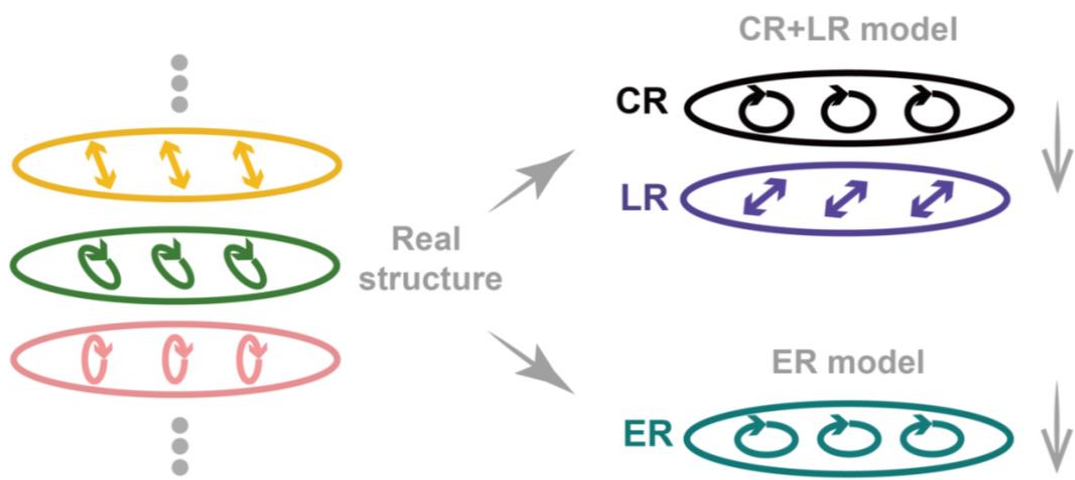

**Fig. 1. Structure of a multi-layer retarder and two separate analysis models.** Here, circles, ellipses, and lines are used to represent different fast axis properties of each 'layer'. The left part shows the general structure of an unknown retarder sample, which may consist of multiple discrete layers or a continuous, non-layered structure that can be approximated as having an infinite number of layers. The upper right part is the pre-defined retarder structure model of the conventional decomposition method: a dual-layer structure with a circular retarder (CR) followed by a linear retarder (LR). This model can sometimes lead to misinterpretation of the polarization information. The lower right part is the physically plausible model: characterizing an unknown sample with multi-layer or diffuse structure as an elliptical retarder (ER). This model provides a more appropriate characterization of the retarder that avoids making any assumptions about the material.

In nature, some retarders consist of discrete layers, while others have no clear layered structure and can be modeled as infinitely layered media. Both cases can be depicted in the left part of **Fig. 1**. For layered retarders, in the case of a two-layer structure, the alignment and the properties of each layer influence the decomposition results, as previously discussed. In the case of a multi-layer structure, both the number of layers and the properties of each layer must be known to construct an accurate model. Simply characterizing a layered retarder using a predefined dual-layer model, as shown on the upper right part of **Fig. 1**, may lead to misinterpretation of its properties. Furthermore, for non-layered retarders, applying a layer-based model is inherently inappropriate. Therefore, we need a more general and meaningful characterization framework to describe such samples. In this context, we find the elliptical retarder representation offers unique advantages for this purpose.

*b) Alternative Vectorial Metrics for Retarder Analysis*

The core of our work is that, for a retarder with an unknown structure, rather than trying to further decompose it into distinct retarder elements, an elliptical retarder model offers a more appropriate characterization, as shown on the lower right part of **Fig. 1**. Regardless of how complex the internal structure of the material may be, its overall effect on polarized light can be described as a single elliptical retarder. Based on this concept, the elliptical axis orientation $\varphi$, degree of ellipticity $\chi$, and elliptical retardance $\rho$ can be used as an effective set of parameters to properly represent the physical properties of a retarder.

The function of a general elliptical retarder can be seen as rotating along a specific axis on the Poincaré sphere by a certain angle. The rotation axis, which is by definition the fast axis of the retarder, corresponds to a certain polarization state $U_R$ on the Poincaré sphere. According to the polarization ellipse, this polarization state can be determined by two parameters, the elliptical axis orientation $\varphi$ (in the range of $[-\frac{\pi}{2}, \frac{\pi}{2}]$) and the degree of ellipticity $\chi$ (in the range of $[-\frac{\pi}{4}, \frac{\pi}{4}]$). The fast axis of the retarder $S_R$ and the corresponding polarization state $U_R$, are given in Eq. (4) [see details in **Supplementary Material 1**]. The parameters $a_1, a_2, a_3$ are the three components of the normalized fast axis vector. Subsequently, $\varphi$ and $\chi$ can be calculated using Eq. (5) and Eq. (6) as follows:

$$S_R = \begin{pmatrix} 1 \\ a_1 \\ a_2 \\ a_3 \end{pmatrix} = \begin{pmatrix} 1 \\ \cos 2\varphi \cos 2\chi \\ \sin 2\varphi \cos 2\chi \\ \sin 2\chi \end{pmatrix} \quad U_R = \begin{pmatrix} \cos 2\varphi \cos 2\chi \\ \sin 2\varphi \cos 2\chi \\ \sin 2\chi \end{pmatrix} (4)$$

$$\varphi = 0.5 \operatorname{atan2}(a_2, a_1) \ (5)$$

$$\chi = 0.5 \sin^{-1} a_3 \ (6)$$

After determining the rotation axis, the elliptical retardance $\rho$ describes the angle of rotation around that axis according to the left-hand rule. The physical meaning of the retardance $\rho$ is the phase shift introduced between the fast axis and slow axis when light propagates through the retarder. The value of $\rho$ can be calculated using Eq. (7) from the bottom right 3×3 submatrix of $M_R$, denoted as $m_R$.

$$\rho = \cos^{-1}\left[\frac{tr(m_R)-1}{2}\right] \ (7)$$

## 3. Results

Three experiments were conducted to demonstrate that the elliptical retarder model provides a more appropriate model for characterizing unknown retarder samples, covering both layered and non-layered structure. First, for samples with a layered structure, two experiments [see details in **Supplementary Material 3**] demonstrated that conventional parameters can lead to misinterpretation of sample properties when the material's actual structure deviates from the predefined model. In addition, the corresponding elliptical retarder model results for each case were also presented. Second, for samples with non-layered structure [see details in **Supplementary Material 4**], where conventional layer-based models are inherently unsuitable, we demonstrate that the proposed physically plausible parameters effectively characterize the polarization properties of the samples. In both experiments, LCs were chosen as retarder samples because of their well-defined and stable birefringent properties. Then, the Mueller matrix of each sample was obtained using an LC-based Mueller matrix polarimeter, and the corresponding parameters were subsequently derived through the MMPD method.

*a) Layer-Structure Sample with alternative alignment*

The procedure for the first experiment on the layer-structure sample was as follows: first, we fabricated the linear and circular retarders and measured the reference linear axis orientation $\theta$ and linear retardance $\delta$ for the linear retarder, and the circular retardance $\phi$ for the circular retarder [see details in **Supplementary Material 3**]. For the fabrication of the linear retarder, we used a nematic LC mixture (E7, Synthon Chemicals Ltd.) capillary filled into a glass cell with anti-parallel rubbed polyimide alignment layers. For the fabrication of the circular retarder, we used a chiral nematic LC mixture consisting of 83 weight percent (wt.%) E7 LC and 17 wt.% R811 chiral dopant (R811, Merck KGaA)[72]. The thickness of the glass cell was 5 microns for the linear retarder and 9 microns for the circular retarder, both of these were determined from the transmission spectrum of a UV-Vis spectrometer. We treat the LC glass cells as a whole, and the effect of the glass on the designed retarders is considered to be negligible.

Second, these two retarders were then combined and arranged in sequence — the linear retarder followed by the circular retarder — forming an overall elliptical retarder ($M_R = M_{CR}M_{LR}$). Third, we measured the Mueller matrix of this elliptical retarder and calculated its parameters according to two decomposition configurations $M_R = M_{LR}M_{CR}$ and $M_R = M_{CR}M_{LR}$, respectively. This allowed us to derive the polarization parameters of the linear retarder and circular retarder for each configuration. Fourth, we compared the derived polarization parameters with the known reference values to see which set of decomposition results accurately characterized the sample properties. Finally, we calculated the same parameters using the elliptical retarder model.

The reference values of the linear and circular retarder, calculated over the entire imaging area and presented in the form *mean value ± standard deviation*, are as follows: for the linear retarder, the linear axis orientation was $\theta = 0.54° \pm 0.41°$ and the linear retardance was $\delta = 144.47° \pm 0.81°$. For the circular retarder, the circular retardance was $\phi = 64.986° \pm 0.30°$. These two samples were then stacked together, with the linear retarder positioned in front and the circular retarder behind, forming a composite elliptical retarder in the experiment, as shown in **Fig. 2a(i)**. We then calculated the results for both decomposition sequences. In the conventional sequence, which is opposite to the actual structure, the linear axis orientation was $\theta = -36.605° \pm 0.342°$, the linear retardance was $\delta = 135.901° \pm 0.777°$, and the circular retardance was $\phi = 70.614° \pm 1.287°$. In the reverse sequence, which aligns with the actual structure, the linear axis orientation was $\theta = -1.298° \pm 0.362°$, the linear retardance was $\delta = 135.901° \pm 0.777°$, and the circular retardance was $\phi = 70.614° \pm 1.287°$. We can see from the results that, in both cases, the linear retardance and circular retardance remain unchanged. However, the linear axis orientation results from the conventional sequence show deviations from the reference values, which may lead to misinterpretation. For example, in certain biomedical imaging, such deviations may result in incorrect interpretation of fiber orientation. In contrast, the reverse sequence, which aligns with the actual structure, yields linear axis orientation results that are close to the reference values [see details in **Supplementary Material 5**]. Additionally, it can be observed that switching from the conventional to the reverse decomposition sequence makes the linear axis orientation shifts from $\theta$ to $\theta + 0.5\phi$, as discussed above. This experiment demonstrated that when the sample's layer sequence differs from the predefined model, conventional parameters may lead to misinterpretations.

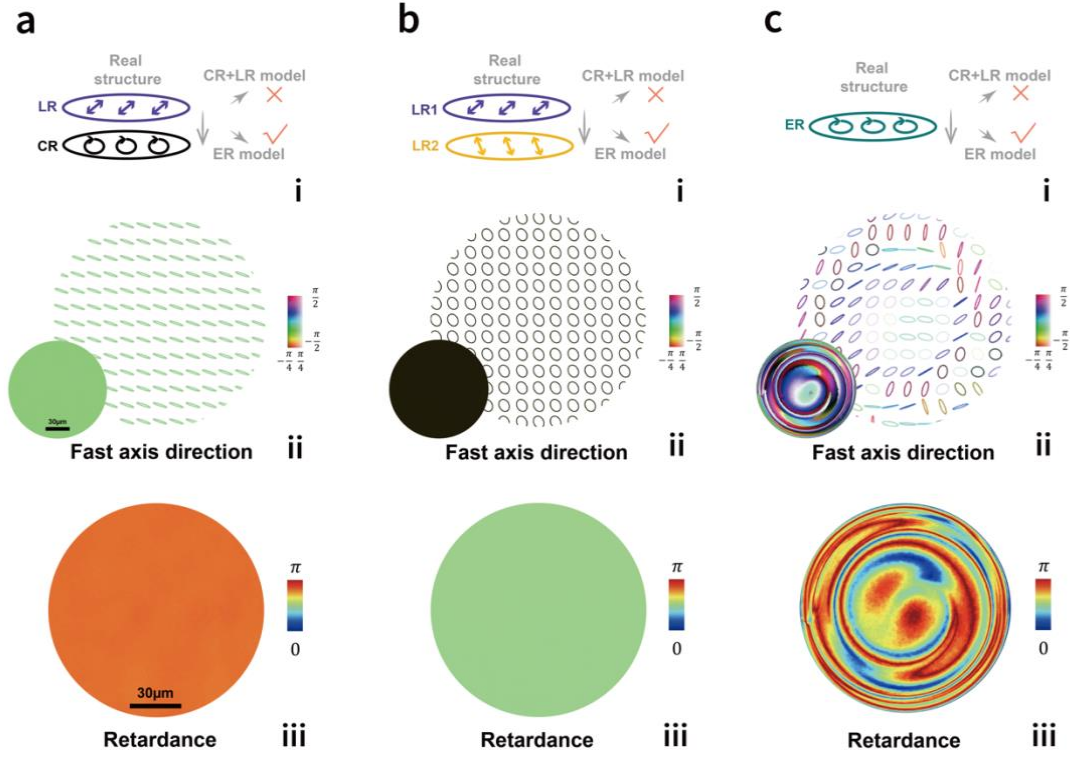

**Fig. 2. Sample structures and ER model characterization**. *a–c,* The configurations of three different LC samples and their characterization using the proposed ER model. *a,* Layered structure composed of a linear retarder followed by a circular retarder, fabricated using a nematic LC and a chiral nematic LC, respectively. *b,* Layered structure composed of two linear retarders with mismatched orientations, both fabricated using nematic LC. *c,* Non-layered structure formed by a chiral nematic LC droplet with spatially varying twist, printed onto a glass substrate with a rubbed polyvinyl alcohol alignment layer using a MicroFab inkjet system. (i) Schematic illustration of the actual sample structure and the different modelling approaches. Conventional models based on a fixed combination of circular and linear retarder often lead to misinterpretation when misaligned with the real structure. In contrast, the ER model provides a physically plausible characterization by capturing the overall polarization effects of the sample. (ii) Fast axis demonstrated using the HSL model, with hue for orientation $\varphi$, luminance for degree of ellipticity $\chi$, and saturation fixed at 1. For illustrative purposes, fast axes were evenly sampled from the imaging region and plotted as ellipses. The colors of the ellipses are assigned in the same way, while their shapes illustrate the polarization states corresponding to each color. (iii) Elliptical retardance distribution of the sample.

Subsequently, we calculated the parameters for the elliptical retarder model, which are visualized in **Fig. 2a(ii)** and **Fig. 2a(iii)**. In the lower-left inset of **Fig. 2a(ii)**, we utilized the Hue-Saturation-Luminance (HSL) model to visualize the fast axis of the sample. Specifically, we used the hue color bar from the model to represent the elliptical axis orientation $\varphi$, luminance to represent the degree of ellipticity $\chi$ (in the lighter regions, the fast axis tends to approach a right-handed circular polarization state, whereas in the darker regions, the fast axis tends to approach a left-handed circular polarization state), and set the saturation value to 1. The corresponding color bar is shown on the right side of **Fig. 2a(ii)**. To make the fast axis more visually intuitive, we sampled points within the region and plotted the fast axes of the ellipses, as shown in the central part of **Fig. 2a(ii)**. The colors of these ellipses are the same as their sampled points, while their shapes represent the polarization states associated with each color. The result suggests that the fast axis of the sample is close to linear, with a low degree of ellipticity. The distribution of the sample's elliptical retardance is illustrated in **Fig, 2a(iii)**. The elliptical retardance across the imaging region was $\rho = 143.830° \pm 1.205°$, which is higher than each of the individual layer. The corresponding subfigures in **Fig. 2b** and **2c** were visualized using the same method.

*b) Layer-Structure Sample with mismatched layer properties*

The procedure for the second experiment on the layered-structure sample was as follows: first, we fabricated two linear retarders and measured its reference linear axis orientation $\theta$ and linear retardance $\delta$ [see details in **Supplementary Material 3**]. The fabrication process of the linear retarders is the same as in the first experiment. Second, these two retarders were combined to form an overall elliptical retarder ($M_R = M_{LR2}M_{LR1}$). Third, we conducted Mueller matrix imaging on this elliptical retarder and calculated its conventional retarder parameters. Next, the derived parameters were compared with the actual configuration to demonstrate how misinterpretation may occur when the actual layer properties differ from those in the predefined model. Finally, the parameters of the elliptical retarder model were calculated and presented.

The reference values of the two retarders, calculated over the entire imaging area and presented, as before, as *mean value ± standard deviation*: for the first linear retarder, the linear axis orientation was $\theta_1 = -14.111° \pm 0.169°$ and the linear retardance was $\delta_1 = 142.969° \pm 0.242°$. For the second linear retarder, the linear axis orientation was $\theta_2 = 86.776° \pm 0.207°$ and the linear retardance was $\delta_2 = 142.277° \pm 0.316°$. These two samples were then stacked to form a composite elliptical retarder for measurement, as shown in **Fig. 2b(i)**. The conventional decomposition of the above configuration produces a circular retarder with high circular retardance $\phi = -84.755° \pm 0.348°$, which is not present in the actual configuration. Additionally, the linear retardance $\delta = 16.035° \pm 0.232°$ is considerably lower than that of the actual configuration. Subsequently, we calculated the parameters for the ER model, which are displayed in **Fig. 2b(ii)** and **Fig. 2b(iii)**. We can see that the retarder sample, composed of two linear retarders with distinct orientations, exhibits a relatively high degree of ellipticity in its fast axis. Additionally, the elliptical retardance across the imaging region was $\rho = 86.451° \pm 0.350°$. The resulting optical properties are similar to those of a quarter-wave plate whose fast axis corresponds to left-handed circular polarization. The above results demonstrate that when the sample's layer properties differ from those in the predefined model, conventional parameters may result in misinterpretations.

It should be noted that the data shows some deviations in these two experiments. We believe that these deviations may be caused by three factors. Firstly, the LC samples we used are not ideal linear or circular retarders. Secondly, the stacked sample thickness is too high to achieve accurate focusing, given that each individual sample is 1.44 mm thick. Thirdly, there is slight spatial non-uniformity in the distribution of the LC. In the measurement process, it is hard to consistently image the exact same region every time. Despite these deviations, we are still able to derive convincing conclusions from the experimental results.

In summary, the two experiments above demonstrate that for a retarder sample with a two-layer structure, conventional parameters may lead to misinterpretations when the material's layer structure differs from the pre-defined model. In the general case, as the number of layers and the properties of each layer in a layered structure are typically unknown in most imaging tasks, it is more reliable to characterize them as elliptical retarders from their overall effects.

*C) Non-Layer-Structure Sample*

The procedure for the nonlayer-structure sample was as follows: first, we fabricated a chiral nematic LC droplet with spatially varying twist as the imaging target. The droplet was printed onto a glass substrate coated with a rubbed polyvinyl alcohol alignment layer using a MicroFab inkjet system [see details in **Supplementary Material 4**]. The LC droplet serves as an example of a non-layered sample, featuring a spatially varying, locally twisted director field that lacks well-defined macroscopic layering. As shown in **Fig. 2c(i)**, this structure differs from a layered configuration of linear and circular retarders, making the conventional dual-layer model inherently unsuitable. Second, we characterized the LC droplets using the physically plausible parameters from the elliptical retarder model. The results validate the feasibility of the proposed parameter set in capturing the polarization properties of non-layered retarders.

From **Fig. 2c(ii)**, it can be observed that the fast axis primarily exists in the form of elliptical polarization states. Specifically, the central region is dominated by right-handed elliptical polarization, while the arrangements in the peripheral regions are more complex. In **Fig. 2c(iii)**, the distribution of the elliptical retardance is shown, which aligns with the fast axis distribution observed in **Fig. 2c(ii)**. Additionally, since each data point corresponds to only one set of fast axes and its elliptical retardance, the overall polarization properties of the sample can be clearly represented. We also conducted imaging of a biomedical slice sample using both conventional and physically plausible retarder parameters as an additional demonstration [see details in **Supplementary Material 6**]. In summary, for non-layer-structure samples, where conventional layer-based models are inherently unsuitable, this experiment demonstrates that physically plausible parameters can clearly characterize the overall polarization properties of the sample.

## 4. Discussion

Due to the non-reciprocity of matrix multiplication and the constraints of predefined models, conventional retarder-related parameters may lead to misinterpretations of sample properties. To avoid this issue, we propose to use a set of physically plausible parameters to characterize a retarder with unknown structure as a whole from the elliptical retarder perspective.

However, further works still need to be done considering elliptical retarder format. To uniquely characterize a retarder, two additional issues need to be considered. Firstly, due to phase wrapping, a $2\pi$ range is assumed in this study, and the retardance values are derived within this range. Secondly, even within a $2\pi$ range (e.g., 0 to $2\pi$), any given retarder has a corresponding retarder with different properties but the same Mueller matrix, and we need to choose one of them. The fast axis of this corresponding retarder is the slow axis of the original retarder, and its retardance is $2\pi$ minus the retardance of the original retarder. These issues arising from the limitation of Stokes-Mueller formalism may be addressed by obtaining more prior knowledge of the sample, or through the introduction of other optical parameters such as wavelength and absolute phase[73,74].

In summary, the physically plausible parameter set provides a new perspective for future research, particularly in the characterization of media with complex structures. By offering a comprehensive characterization of a retarder, this work lays the foundation for potential applications in areas such as biomedical analysis and material characterization.

**Supplementary Material**

See the supplementary material for additional details on the interpretation of the retarder, intuitive understanding of the parameter's relationship, fabrication and measurement of retarder samples, comparison of measurement results, and additional imaging results.

**Acknowledgments**

The authors would like to thank the support of St John's College, the University of Oxford, and the Royal Society (URF\R1\241734) (C.H.). The authors would also like to thank Prof. Martin Booth and Prof. Daniel Royston at the University of Oxford for their valuable support.

**Author Declarations**

*Conflict of interest*

The authors have no conflicts to disclose.

*Author Contributions*

**Runchen Zhang**: Conceptualization (equal); Formal analysis (equal); Investigation (lead); Visualization (equal); Methodology (equal); Writing – original draft (lead); Writing – review & editing (equal). **Xuke Qiu**: Investigation (supporting); Visualization (equal); Methodology (equal). **Yifei Ma**: Investigation (supporting); Visualization (equal); Methodology (equal). **Zimo Zhao**: Investigation (supporting); Visualization (equal); Methodology (equal). **An Aloysius Wang**: Formal analysis (equal); Visualization (supporting); Methodology (equal). **Jinge Guo**: Investigation (supporting); Visualization (equal); **Ji Qin**: Investigation (supporting); Visualization (equal). **Steve J. Elston**: Formal analysis (supporting);

Visualization (supporting). **Stephen M. Morris**: Supervision (equal); Resources (supporting); Project Administration (equal); Writing – review & editing (equal). **Chao He**: Conceptualization (equal); Supervision (equal); Project Administration (equal); Resources (lead); Writing – review & editing (lead).

**Data Availability**

The data that support the findings of this study are available from the corresponding author upon reasonable request.